\documentstyle[aps]{revtex}

\begin{document}
\author{Daxiu Wei\thanks{%
Electronic mail: daxiuwei@wipm.whcnc.ac.cn }, Xiaodong Yang, Jun Luo,
Xianping Sun, Xizhi Zeng, Maili Liu, and Shangwu Ding}
\title{Experimental realization of 7-qubit universal perfect controlled-NOT and
controlled square-root NOT gates }
\address{Laboratory of Magnetic Resonance and Atomic and Molecular Physics, Wuhan\\
Institute of Physics and Mathematics, The Chinese Academy of Sciences, \\
Wuhan 430071, People's Republic of China}
\maketitle

\begin{abstract}
The controlled-NOT gate and controlled square-root NOT gate play an
important role in quantum algorithm. This article reports the experimental
results of these two universal quantum logic gates (controlled square-root
NOT gate and controlled-NOT gate) on a 7-qubit NMR quantum computer.
Further, we propose a simple experimental method to measure and correct the
error in the controlled phase-shift gate, which is helpful to construct a
more perfect phase-shift gate experimentally and can also be used in more
qubits discrete Fourier transformation.\newline
\newline
PACS number(s): 03.67.Lx. 76.60.-k. 75.10.Jm
\end{abstract}

\section{Introduction}

By using the characteristics of quantum mechanics, quantum computers (QC)
are faster than classical computers when performing certain computations
such as the factorization of a large number \cite{shor}, searching of
database \cite{grover} especially when simulating quantum systems \cite
{lloyd}. Among many schemes of realizing quantum computers (for example
trapped ions \cite{cirac}, cavity QED \cite{turchette} , quantum dots \cite
{band}, NMR \cite{chuang,cory}), the scheme based on liquid NMR techniques
has made remarkable progress \cite{jones}. In experiments, the four-qubit
entanglement has been implemented by using trapped ions \cite{sacette}, and
seven-qubit cat-state \cite{knill}, five-qubit D-J algorithm \cite{marx},
order finding algorithm \cite{lieve} by using liquid NMR techniques. People
are most interested in extension quantum computation to multi-qubit spin
systems \cite{berman,cirac1}.

However, it is a technical challenge to extend to more qubits experimentally
due to the low signal-to-noise (S/N) ratio of NMR and further, the S/N ratio
will decrease exponentially with the increasing of qubit \cite{warren}.
Hence, the quantum computing based on liquid NMR will reach no more than 10
qubits \cite{warren}. With the increasing of the number of qubit, it is not
easy to selectively control the coherent evolution between two specific
spins as the coupling network becomes more complex. Though, efficient
methods have been developed to refocus the interaction of single-spin and
two spins \cite{jones1}. In a multi-qubit NMR QC, it is necessary to
suppress the interactions coming from the other qubits when realizing the
quantum logic gate between two given spins \cite{jones1}. Moreover, in the
process of achieving complicated quantum computation, the precision of every
gate plays a very important role in realizing quantum computing \cite{knill1}%
. It is a big technical challenge if we use homonuclear system to realize
multi-qubit computing, due to the use of selective pulse sequences. So, it
is necessary to establish a set of measurement to examine and correct the
quantum logic gate.

The controlled-NOT (CNOT) gate and controlled square-root NOT gate \cite
{jones} play an central role in many quantum computations (especially
quantum Fourier Transformation \cite{copper}, the Grover's searching
algorithm \cite{grover}, the Toffoli gate \cite{divincenzo}). In this
article, we report the experimental realization of controlled-NOT gate and
controlled square-root NOT gate between two given qubits on seven-qubit NMR
quantum computers, and we also propose a method to verify and correct the
phase-shift error.

\section{Perfect controlled-NOT gate and controlled square-root NOT gate
(universal phase-shift gate)}

CNOT gate is a kind of universal controlled phase-shift gate, which is
defined by \cite{jones}:

\ 
\begin{equation}
\mid 1\rangle \mid 1\rangle \stackrel{\varphi }{\longrightarrow }e^{i\varphi
}\mid 1\rangle \mid 1\rangle .
\end{equation}

The quantum circuit to realize a controlled square-root NOT gate is shown in
figure 1. h and h $^{-1}$are pseudo-Hadamard gates, which can be implemented
by 90 degree pulses along y axis. The phase-shift gate $Z$ $_{\Phi }$ of an
arbitrary angle can be realized by selective composite z-pulses:

\begin{equation}
\frac{1}{4nJ_{ij}}-180_{x}^{\circ }-\frac{1}{4nJ_{ij}}-90_{x}^{\circ }-(%
\frac{\varphi }{2})_{-y}-90_{-x}^{\circ }.\newline
\end{equation}
Pulses here affects two given spins i and j simultaneously. When n=1, the
logic gate according to figure.1 is a controlled-NOT gate. When n=2, it is a
controlled square-root NOT gate. When n is an arbitrary integer, it is a
controlled n-th square-root NOT gate.

We select a seven-qubit spin system ( $^{13}$C-labeled trans-crotonic acid (C%
$^{1}$H$_{3}^{3}$ C$^{2}$H$^{1}$=C$^{3}$H$^{2}$C$^{4}$O$_{2}$H) \cite{knill}
), and begin the experiment from a thermal equilibrium state in a product
operator representation, we then measure the phase-shift $\Phi $ and correct
it as follow, which is simple and feasible in the experiment:

\begin{eqnarray}
&&\mu _{C}(I_{z}^{C^{1}}+I_{z}^{C^{2}}+I_{z}^{C^{3}}+I_{z}^{C^{4}})+\mu
_{H}(I_{z}^{H^{1}}+I_{z}^{H^{2}}+I_{z}^{H^{3}})  \nonumber \\
&&\stackrel{(\pi /2)_{-y}^{C^{2}}}{\longrightarrow }\mu
_{C}(I_{z}^{C^{1}}+I_{x}^{C^{2}}+I_{z}^{C^{3}}+I_{z}^{C^{4}})+\mu
_{H}(I_{z}^{H^{1}}+I_{z}^{H^{2}}+I_{z}^{H^{3}})  \nonumber \\
&&\stackrel{1/2nJ_{C^{1}C^{2}}}{\longrightarrow }\mu
_{C}(I_{z}^{C^{1}}+I_{x}^{C^{2}}\cos \frac{\pi }{2n}%
-2I_{y}^{C^{2}}I_{z}^{C^{1}}\sin \frac{\pi }{2n}%
+I_{z}^{C^{3}}+I_{z}^{C^{4}})+\mu
_{H}(I_{z}^{H^{1}}+I_{z}^{H^{2}}+I_{z}^{H^{3}})  \nonumber \\
&&\stackrel{(\pi /2)_{X}^{C^{2}}}{\longrightarrow }\mu
_{C}(I_{z}^{C^{1}}+I_{x}^{C^{2}}\cos \frac{\pi }{2n}%
+2I_{z}^{C^{2}}I_{z}^{C^{1}}\sin \frac{\pi }{2n}%
+I_{z}^{C^{3}}+I_{z}^{C^{4}})+\mu
_{H}(I_{z}^{H^{1}}+I_{z}^{H^{2}}+I_{z}^{H^{3}})  \nonumber \\
&&\stackrel{PFG}{\longrightarrow }\mu
_{C}(I_{z}^{C^{1}}+2I_{z}^{C^{2}}I_{z}^{C^{1}}\sin \frac{\pi }{2n}%
+I_{z}^{C^{3}}+I_{z}^{C^{4}})+\mu
_{H}(I_{z}^{H^{1}}+I_{z}^{H^{2}}+I_{z}^{H^{3}})  \nonumber \\
&&\stackrel{(\pi /2)_{X}^{C^{2}}}{\longrightarrow }\mu
_{C}(I_{z}^{C^{1}}+2I_{y}^{C^{2}}I_{z}^{C^{1}}\sin \frac{\pi }{2n}%
+I_{z}^{C^{3}}+I_{z}^{C^{4}})+\mu
_{H}(I_{z}^{H^{1}}+I_{z}^{H^{2}}+I_{z}^{H^{3}})  \nonumber \\
&&\stackrel{(\Phi )_{z}^{C^{2}}}{\longrightarrow }\mu
_{C}(I_{z}^{C^{1}}+2I_{y}^{C^{2}}I_{z}^{C^{1}}\sin \frac{\pi }{2n}\cos \Phi
+2I_{x}^{C^{2}}I_{z}^{C^{1}}\sin \frac{\pi }{2n}\sin \Phi
+I_{z}^{C^{3}}+I_{z}^{C^{4}})+\mu
_{H}(I_{z}^{H^{1}}+I_{z}^{H^{2}}+I_{z}^{H^{3}}),\newline
\end{eqnarray}
C $^{2}$ here denotes selective pulses on Carbon 2. 1/2nJ$_{C^{1}C^{2}}$ is
J-coupling evolution. $\mu _{C}$ and $\mu _{H}$ are the nuclear magnetic
moments. PFG denotes the pulsed field of gradient. $\left( \Phi \right)
_{z}^{C^{2}}$expresses selective composite z-pulses on C$^{2}$ , and the
refocusing and decoupling pulses are not given here.

We have used the following two steps to measure the phase shift $\Phi $ :

(1). Add a selective read-pulse ($\frac{\pi }{2}$)$_{x}^{C^{2}}$ behind the
above pulses, the product operator is :

\begin{equation}
\mu _{C}(I_{z}^{C^{1}}-2I_{z}^{C^{2}}I_{z}^{C^{1}}\sin \frac{\pi }{2n}\cos
\Phi +2I_{x}^{C^{2}}I_{z}^{C^{1}}\sin \frac{\pi }{2n}\sin \Phi
+I_{z}^{C^{3}}+I_{z}^{C^{4}})+\mu
_{H}(I_{z}^{H^{1}}+I_{z}^{H^{2}}+I_{z}^{H^{3}}),\newline
\end{equation}
here the signal is $2$I$_{x}^{C^{2}}$I$_{z}^{C^{1}}\sin \frac{\pi }{2n}\sin
\Phi $.

(2). Add a selective read-pulse ($\frac{\pi }{2}$)$_{y}^{C^{2}}$behind the
above pulses, the product operator is :

\begin{equation}
\mu _{C}(I_{z}^{C^{1}}+2I_{y}^{C^{2}}I_{z}^{C^{1}}\sin \frac{\pi }{2n}\cos
\Phi +2I_{z}^{C^{2}}I_{z}^{C^{1}}\sin \frac{\pi }{2n}\sin \Phi
+I_{z}^{C^{3}}+I_{z}^{C^{4}})+\mu
_{H}(I_{z}^{H^{1}}+I_{z}^{H^{2}}+I_{z}^{H^{3}}),\newline
\end{equation}
here the signal is $2$I$_{y}^{C^{2}}$I$_{z}^{C^{1}}\sin \frac{\pi }{2n}\cos
\Phi $.

By repeatedly adjusting the phases of the receiver, we obtain the integrated
intensities of the two experimental signals. The phase-shift $\Phi $ can be
calculated from the ratio $\tan \Phi $. Then the $\Phi $ of the composite
pulses and the evolution time are adjusted to correct the error so as to
obtain a more accurate controlled square-root NOT gate.

\section{Experiment and Result}

We carried out the experiment with a Bruker-ARX500 spectrometer. The sample
was 20mg crotonic acid (U-$^{13}$C$_{4\text{,}}$ Cambridge Isotope
Laboratories Inc. Cat. No. CLM-6118) dissolved in deuterated acetone, which
was degassed and flame-sealed in a standard 5mm NMR test tube. The
experimental temperature was maintained at 25$^{\circ }$C. The molecular
structure of crotonic acid can be found in \cite{knill}. We have used four
carbons and three protons as the seven qubits. The shape of the soft pulse
is Gaussian. Typical experiment results are shown in figures 2 and 3. Both
are C$^{2}$ NMR spectra (not decoupled). The theoretical and experimental
results of $\Phi $ in the controlled square-root NOT gate, and the relative
errors $\mid \frac{\Phi _{\exp }-\Phi _{\text{the}}}{\Phi _{\text{the}}}$ $%
\mid $ between the theoretical phase-shift $\Phi _{\text{the }}$and
experimental $\Phi _{\exp }$ are shown in table 1. The largest error is
2.9\%. Using the method in the last section, where we have adjusted the
parameters of the Gaussian pulses and the evolution time, we achieved to
control the precision of the controlled-NOT gate and realized a more perfect
conditional-phase-shift gate. The experimental errors mainly come from: (1)
the inhomogeneity of the static and radio-frequency magnetic field, (2) the
imperfection of the selective pulse, (3) the inaccuracy of the 90 degree
pulse, and (4) the decaying of signal during the acquiring time.

\section{Conclusion}

We realized the controlled-NOT and controlled square-root NOT gates between
two given qubits on a seven-qubit NMR quantum computer. The experimental
result shows that the control of the other qubit's evolution is successful.
We also give a method to improve the accuracy of the phase-shift gate, which
can be applied to many quantum algorithms, such as multi-qubit Fourier
Transformation. Besides, it is also possible to implement this, the method
of measuring and correcting the error of the controlled square-root NOT gate
as discussed before, into the design of other logic gates in a multi-qubit
quantum computer.

\section{Acknowledgments\newline
}

This work has been supported by the National Natural Science Foundation of
China (Grant No.19874073), the National Knowledge Innovation Program of the
Chinese Academy of Sciences (Grant No. KJCX2-W1). We thank Dr. Xi-an Mao,
and Ms. Hanzhen Yuan for valuable discussions and technical assistance.

\newpage

Table 1. The comparison between $\Phi _{\text{the }}$and $\Phi _{\exp }$.

\begin{tabular}{cccccc}
\hline\hline
$\Phi _{\text{the}}$ & $2I_{x}^{C^{2}}I_{z}^{C^{1}}\sin \frac{\pi }{2n}\sin
\Phi $ & $2I_{y}^{C^{2}}I_{z}^{C^{1}}\sin \frac{\pi }{2n}\cos \Phi $ & $\tan
\Phi $ & $\Phi _{\text{exp}}$ & $\mid \frac{\Phi _{\exp }-\Phi _{\text{the}}%
}{\Phi _{\text{the}}}\mid (\%)$ \\ \hline\hline
90$^{\circ }$ & 1.00\  & 0.02 & 50 & 88.9$^{\circ }$ & 1.3 \\ 
180$^{\circ }$ & 0.09 & -1.00 & -0.09 & 174.9$^{\circ }$ & 2.9 \\ 
270$^{\circ }$ & -1.00 & 0.05 & 20 & 267.1$^{\circ }$ & 1.1 \\ 
360$^{\circ }$ & 0.07 & 1.00 & 0.07 & 356.0$^{\circ }$ & 1.1 \\ \hline\hline
\end{tabular}

\bigskip {\bf Figure captions}

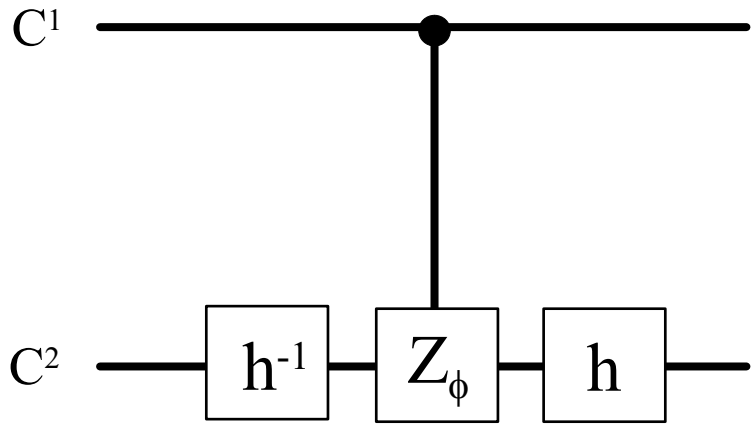
Fig. 1: A quantum circuit for implementing a controlled square-root NOT
gate, C$^{1}$ the controlled qubit and C$^{2}$ the target qubit.

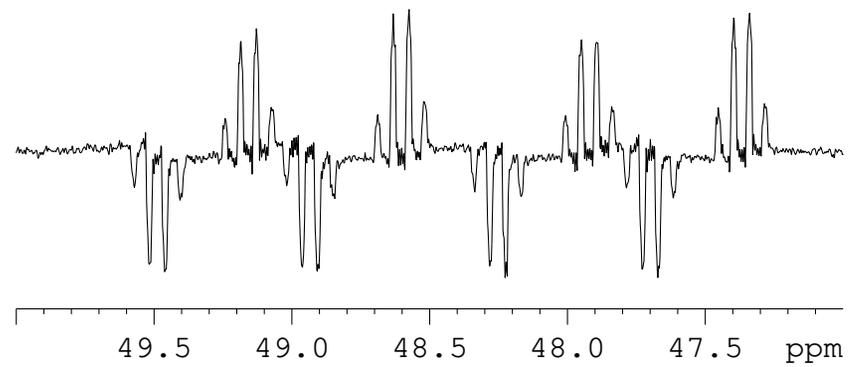
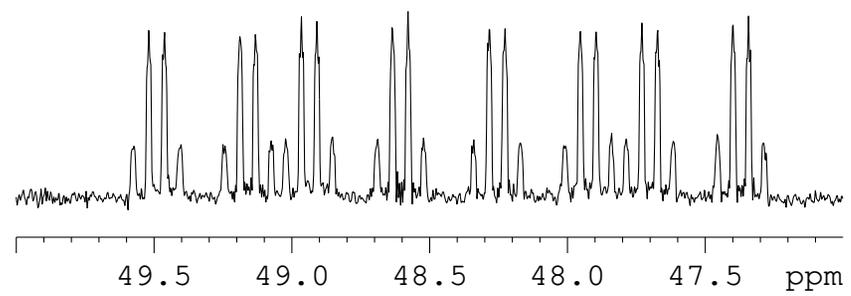
Fig. 2: The spectra of C$^{2}$. Both spectra were acquired with 8 scans.
Results of realizing a controlled-NOT gate (a) and the thermal equilibrium
state (b). The experimental results are in good agreement with the
theoretical.

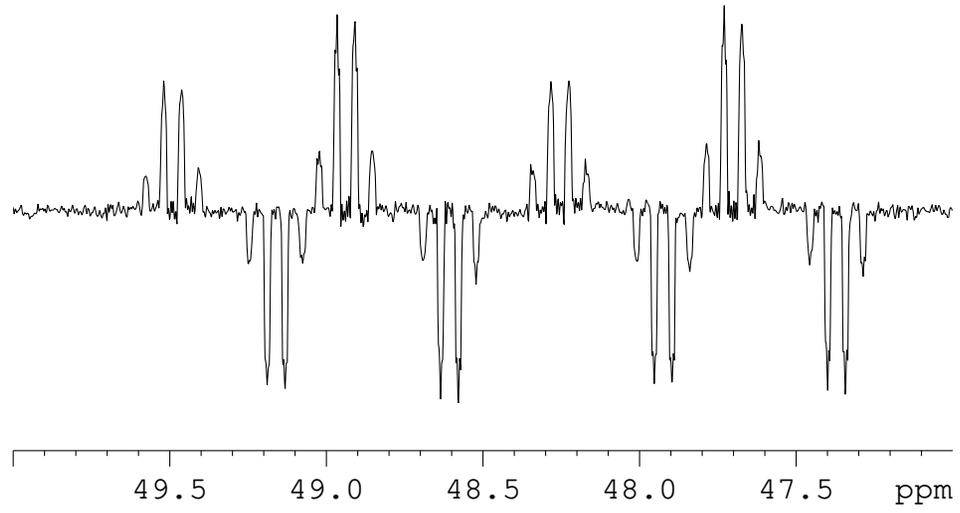
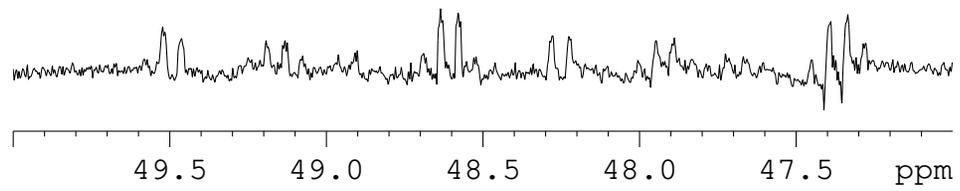
Fig. 3: The spectra of realizing a controlled square-root NOT gate. Both
spectra were acquired with 8 scans. Signals of $2$I$_{x}^{C^{2}}$I$%
_{z}^{C^{1}}\sin \frac{\pi }{2n}\sin 270^{\circ }$ (a) and $%
2I_{y}^{C^{2}}I_{z}^{C^{1}}\sin \frac{\pi }{2n}\cos 270^{\circ }$ (b).

\end{document}